\documentclass[superscriptaddress]{revtex4-1}
\usepackage{amssymb}
\usepackage{amsmath}
\usepackage{graphicx}
\usepackage[normalem]{ulem}
\usepackage[dvips]{color}
\usepackage{array,longtable,multirow}
\usepackage{latexsym}
\usepackage{braket}
\begin{document}
\title{Quantum Zeno dynamics of a matter-wave bright soliton}

\author{Xin Zhang}
\affiliation{Department of Physics, Nanjing University,
Nanjing 210008, China}

\author{Xinwei Fan}
\affiliation{Department of Physics, Nanjing University,
Nanjing 210008, China}

\author{Chang Xu \footnote{Corresponding author}\footnote{cxu$@$nju.edu.cn}}
\affiliation{Department of Physics, Nanjing University,
Nanjing 210008, China}

\author{Zhongzhou Ren \footnote{Corresponding author}\footnote{zren$@$nju.edu.cn, zren@tongji.edu.cn}}
\affiliation{Department of Physics, Nanjing University,
Nanjing 210008, China}
\affiliation{School of Physics Science and Engineering, Tongji University, Shanghai 200092, China}

\author{Jie Peng}
\affiliation{Laboratory for Quantum Engineering and Micro-Nano Energy Technology
and School of Physics and Optoelectronics, Xiangtan University, Hunan 411105,
People's Republic of China}

\begin{abstract}
The quantum measurement problem, namely how the deterministic quantum evolution leads to probabilistic measurement outcomes, remains a profound question to be answered. In the present work, we propose a spectacular demonstration and test of the subtle and peculiar character of the quantum measurement process. We show that a bright soliton supported by a Bose-Einstein condensate can be reflected as a whole by an electron beam, with neither attraction nor repulsion between the condensate's neutral atoms and the beam's electrons. This macroscopic reflection is purely due to the quantum Zeno dynamics induced by the frequent position measurement of the condensate's atoms by the electron beam. As an example of application, just as a soccer player would stop a coming ball, an electron beam moving backward with half the velocity of the bright soliton can precisely stop the soliton. This offers an entirely new and useful tool for manipulating bright solitons.
\end{abstract}

\maketitle

\clearpage
\newpage

If we assume-according to our current best understanding-that the evolution
of the world is governed by the quantum theory, how then can this underlying
deterministic evolution enable the quantum measurement process to give
probabilistic outcomes?
This quantum measurement problem remains
a profound question to be answered \cite{bru,leg}.
In particular, due to the peculiar laws of quantum mechanics, frequently ascertaining
whether a system is inside a region can effectively set up
an impenetrable wall around this region. This is called the quantum Zeno dynamics
(QZD) \cite{fac2002,fac2008}. It is deeply related
to the quantum measurement
process \cite{hom} and thus studies of it can help unveil the secrets behind
the hitherto mysterious measurement process.
QZD has been realized in various physical systems
\cite{sch,sig,bar,bre,kal}.
Besides its
fundamental significance, it has been realized to be a useful
tool for quantum state engineering \cite{kal,wan,man,ros,sha,cha,shi,sha2017}.
Interestingly however,
so far experimental realizations of the QZD have been mainly focused
on systems composed of several to a
handful of discrete quantum levels,
while the QZD of the continuous spatial distribution
of a quantum mechanical system has been
much less explored \cite{bar2013,bra2009,zez2012}.
Recently, such spatial QZD has been observed in
a Bose-Einstein condensate depleted by an electron beam \cite{bar2013}.
The electron beam can be seen as a continuous position measurement
constantly ascertaining whether any atom is in the region irradiated by the beam.
It was observed that beyond some threshold,
a more intense electron beam led to less depletion of the
condensate, evidencing that the atoms in the condensate
are being repelled by the stronger measurement, i.e., the QZD.

\begin{figure*}[htb]
\centering
\includegraphics[width=16.0cm]{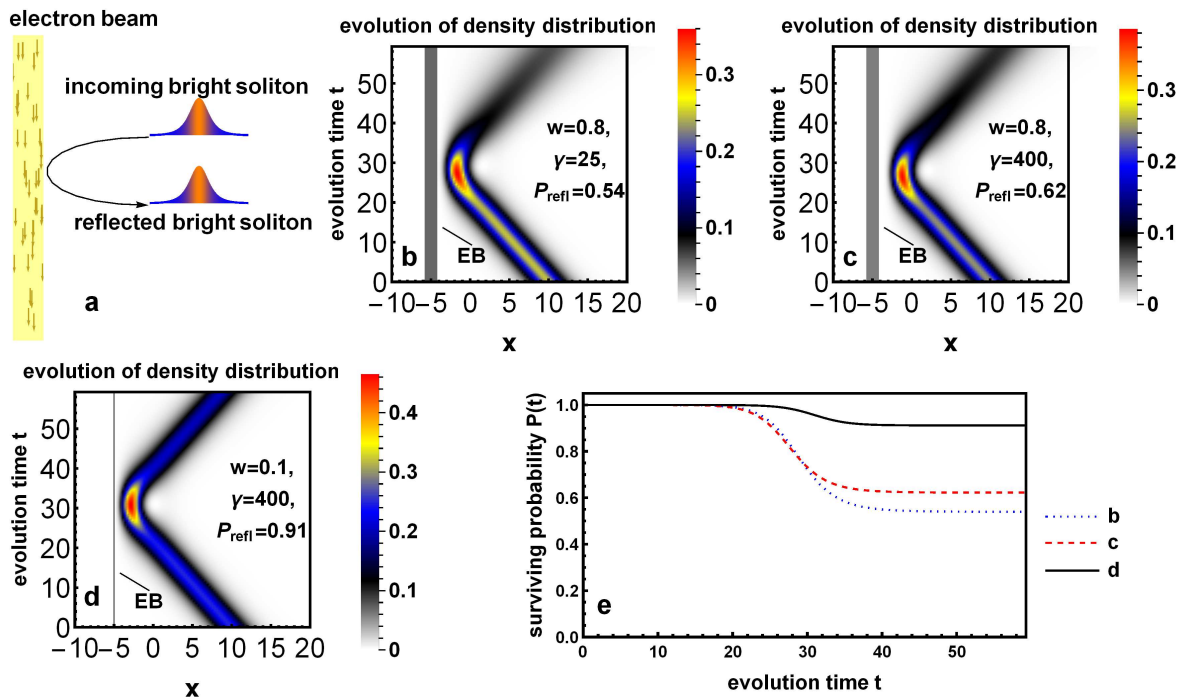}
\caption{Concept and core messages.
(a) An electron beam can reflect a matter-wave bright soliton throgh
QZD. The bright soliton is formed from a Bose-Einstein condensate
with attractive interaction. Normally, an electron
beam simply knocks atoms present in its beam path out of the
condensate. This can also be viewed as ascertaining whether any
atom is along the beam path. When the electron beam is
intense enough, this frequent ascertaining induces a QZD that repel
atoms from entering the beam path. We show that the matter-wave
bright soliton will also be reflected by this QZD repulsion,
and that it can remain a soliton after being reflected.
(b) Calculated time evolution of the density distribution of
an incoming bright soliton. The horizontal axis is the
position, while the vertical axis is the evolution time.
The gray rectangular region (denoted ``EB")
represents the electron beam.
The soliton is reflected by the electron beam.
After being reflected, the solitonic character is not
preserved and the wave packet spreads out.
(c) Same as (b) but for a stronger dissipation strength $\gamma$.
The reflected fraction $P_{\textrm{refl}}$ rises from 0.54 to
0.62. The wave packet after reflection spreads out slower.
(d) Same as (c) but for a sharper beam edge, $w=0.1$ compared
to $w=0.8$ in (c). The reflected
fraction further rises from 0.62 to
0.91. Most importantly, the solintonic character is
well preserved by the reflection in that the wave packet
after reflection propagates with a constant shape.
(e) The surviving fraction of the atoms as a
function of time. The three curves correspond to subfigures
(b)-(d) respectively.
}\label{fig1}
\end{figure*}

\begin{figure*}[htb]
\centering
\includegraphics[width=18.0cm]{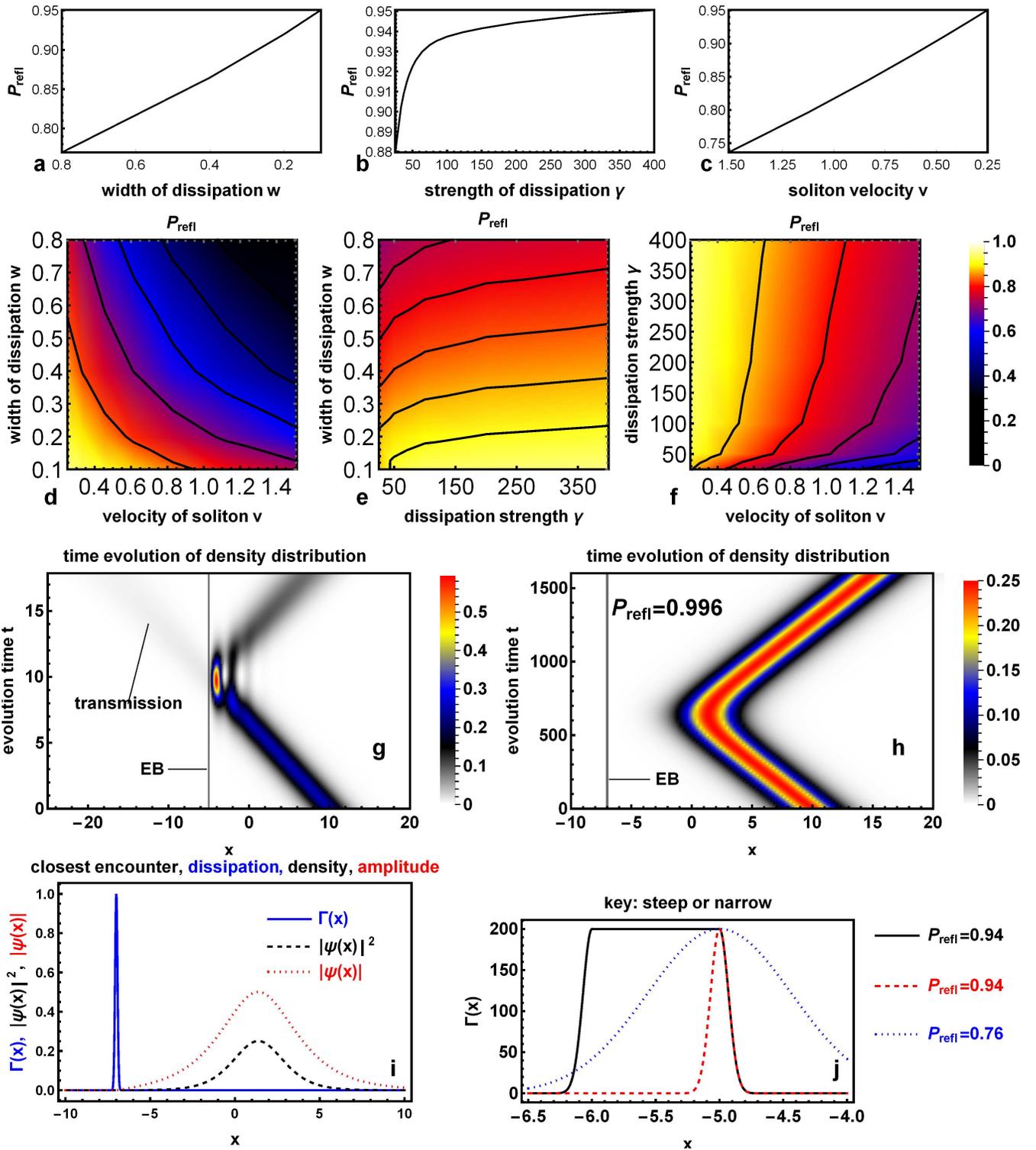}
\caption{Detailed analysis of the QZD reflection of a bright
soliton.
(a) The reflected fraction $P_{\textrm{refl}}$
as a function of the sharpness parameter $w$ of the electron beam.
As can be seen, $P_{\textrm{refl}}$ increases
as $w$ decreases, namely, for a sharper beam edge.
(b)$P_{\textrm{refl}}$ increases with the strength parameter $\gamma$
of the dissipation.
(c) $P_{\textrm{refl}}$  increases as the incoming velocity
$v$ of the bright soliton decreases. (d) Contour plot
of $P_{\textrm{refl}}$ as a function of both $w$ and $v$.
$P_{\textrm{refl}}$ increases as $w$ or $v$ decreases.
(e) $P_{\textrm{refl}}$ increases
as $w$ decreases or $\gamma$ increases. (f) $P_{\textrm{refl}}$
increases as $v$ decreases or $\gamma$ increases.
(g) In a small minority of cases,
if the incoming velocity is high, the
dissipation strength is small
and the beam is narrow, a faint transmission is visible.
(h) For very small incident velocity, the reflection
$P_{\textrm{refl}}$ can be very high.
Curiously, in such cases
it is very pronounced that the soliton appears to be
repelled long before
it even hits the beam. This may seem impossible.
However, as shown in (i), at the closest encounter (t=640),
though the density $|\psi(x)|^2$ seems quite separated
from the beam, the amplitude $|\psi(x)|$ has a sizable presence
near the beam and is able to interact with the beam and effect
the reflection.
(j) The key for effective reflection is a sharp edge but not
narrow beam width. A wide beam with a sharp edge (black solid)
reflects as efficiently as a narrow sharp beam (red dashed),
both achieving $P_{\textrm{refl}}=0.94$.
}\label{fig2}
\end{figure*}

\begin{figure*}[htb]
\centering
\includegraphics[width=16.0cm]{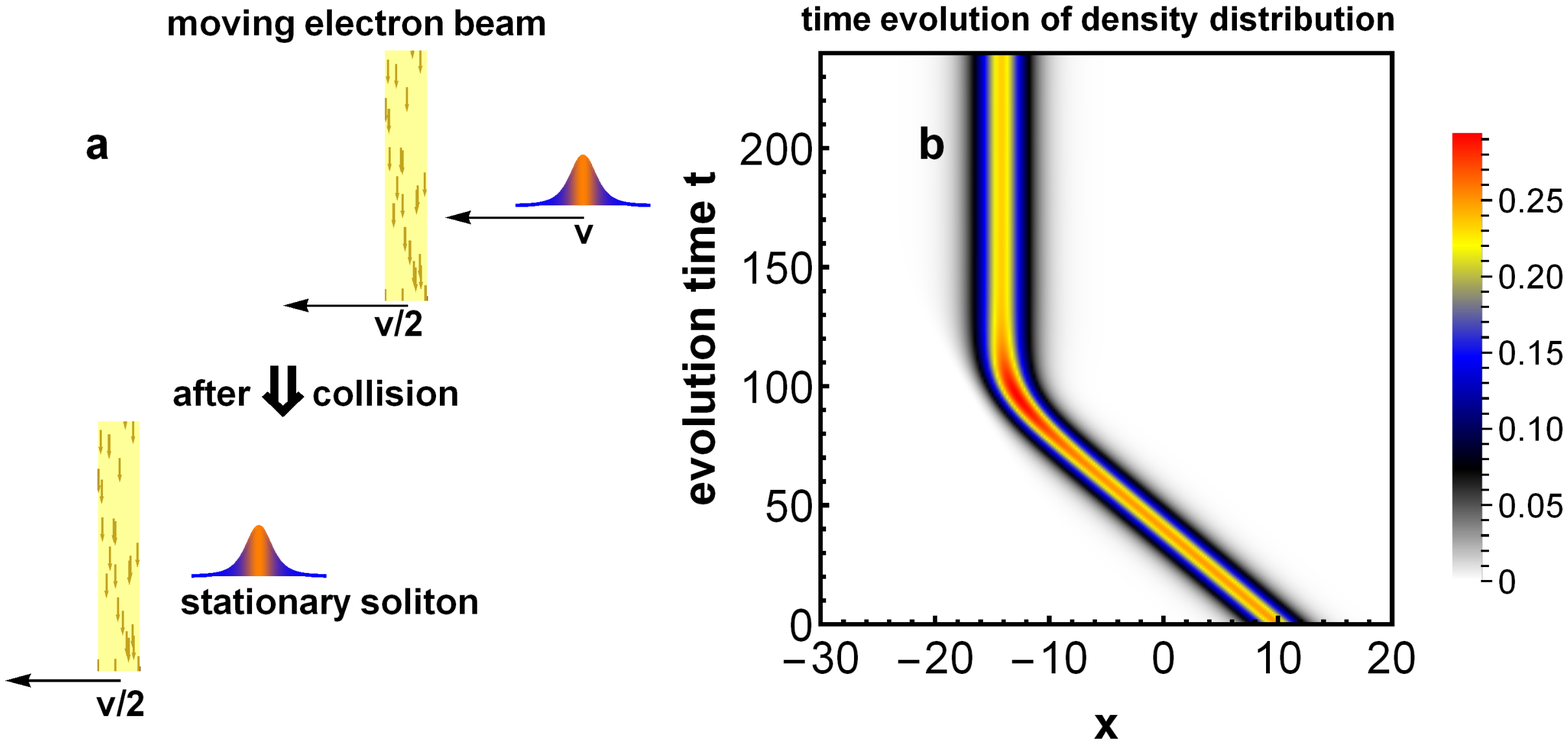}
\caption{
(a) An example of application:
an electron beam moving backward with half the velocity
of the incoming soliton can precisely stop the soliton.
(b) Calculated time evolution of the density distribution.
After encounter with the electron
beam at around $t\sim100$ the soliton's translational motion is precisely stopped.
}\label{fig3}
\end{figure*}

In the present work, we show a spectacular demonstration
and test of the subtle and peculiar character of the quantum  measurement process
can be realized using this new experimental capability
and a bright soliton via QZD.

Solitons is one of the most fascinating phenomena in Bose-Einsten condensates.
Unlike ordinary waves that constantly changes
their shape and often very soon spreads out and disappears,
a soliton is a solitary wave packet that retains its shape
as it propagates, due to a cancellation between dispersion
and nonlinearity.
Studies of solitons have been widely carried out in the
context of Bose-Einstein condensates \cite{gli,ped,str,den,bur,bec,and,eie,kha,ngu,mey}
and helps to reveal
the properties of this exotic state of matter.
In particular, a Bose-Einstein condensate with attractive
atom-atom interactions can support a bright soliton with
a localized density distribution \cite{kha},
which holds great promise
for application in precision
interferometry \cite{mcd,hel,ger,hel2012}.
It is known that the behavior of a bright soliton is akin
to a classical particle. In particular,
just as a classical particle, it can be
repelled by a potential barrier and preserve its solitonic
character \cite{mar}.

In the present work we show that a bright soliton
can be repelled as a whole by the electron beam.
Since there is neither attraction nor repulsion between the condensate's atoms
and the beam's electrons, this spectacular reflection at the macroscopic scale
is purely due to the QZD induced by
the frequent position measurement of
the condensate's atoms by
the electron beam, and thus purely attributable to the peculiar
character of the quantum measurement process.

We show that the key for the QZD induced by
an electron beam to repel a bright soliton and preserve
its solitonic character,
and at the same time cause as small atom loss
as possible,
is to have a sharp beam edge,
apart from having a high enough intensity
as one would probably naturally expect for QZD. That is,
the intensity of the electron beam should rise
from zero to a high value in a short enough distance.
Also, if one has the freedom to choose to work with
a slower soliton, the performance would also improve dramatically.

As an example of application of the proposed phenomenon,
we show that just as a soccer player
would stop a coming ball, a backward moving electron beam with
half the velocity of the coming bright soliton can precisely
stop the translational motion of the soliton.
This offers an entirely new and useful tool for manipulating bright solitons.

\textbf{Results}

\textbf{Physical system}.
We consider a one-dimensional system: A bright soliton
supported by an attractive Bose-Einstein condensate
moves toward an electron beam (cf. Fig. \ref{fig1}a).
We follow the time evolution of the
system
by solving the time-dependent Gross-Pitaevskii equation with a
dissipation
term \cite{bar2013,bra2009,zez2012}:
\begin{eqnarray}\label{tdGP}
i \hbar \frac{\partial \psi(x,t)}{\partial t}=- \frac{\hbar^2}{2 m}
\frac{\partial^2}{\partial x^2}\psi(x,t)- g |\psi(x,t)|^2 \psi(x,t)
-i \gamma \Gamma(x) \psi(x,t).
\end{eqnarray}
Here $\psi(x,t)$ is the wave function of the
condensate.
$g>0$ is the nonlinearity parameter arising from
attractive atom-atom interaction.
The electron beam knocks atoms present in its beam path out
of the condensate. This dissipative effect is
characterized by the strength parameter $\gamma$
and the beam profile
\begin{eqnarray}\label{Gaussianbeam}
\Gamma(x)=e^{-(\frac{x-x_\textrm{b}}{w})^2},
\end{eqnarray}
where $w$ characterizes the sharpness as well as the width of
the electron beam while $x_\textrm{b}$ is the
position of the center of the beam.
Without loss of generality, we set $m=\hbar=g=1$, and normalize
the initial $\psi(x,0)$
such that
\begin{eqnarray}\label{normal}
\int_{-\infty}^{\infty} dx |\psi(x,0)|^2=1.
\end{eqnarray}

\textbf{Quantum Zeno dynamics of a matter-wave bright soliton.}
In spatial regions with negligible dissipation,
equation (\ref{tdGP}) supports the
bright soliton solution:
\begin{eqnarray}\label{soliton}
\psi_{\textrm{soliton}}(x,t)=
A e^{i v x} \textrm{Sech}\left[A \sqrt{g} (x-x_0-v t) \right]
e^{-\frac{1}{2} i t \left(v^2-A^2 g\right)},
\end{eqnarray}
which describes a solitary wave that at $t=0$
is localized around $x_0$ and
moves with velocity $v$ while maintaining its shape unchanged.
$A$ is the amplitude of the soliton.
For our initial condition, we set $t=0$ in the above equation
and normalize
according to equation (\ref{normal}). The initial condition reads:
\begin{eqnarray}
\psi_{\textrm{ini}}(x)=
\frac{\sqrt{g}}{2} e^{i v x}
\textrm{Sech}\left[\frac{g}{2} (x-x_0-v t) \right].
\end{eqnarray}

We start the bright soliton in
a region with negligible dissipation toward the
electron beam (cf. Fig. \ref{fig1}a) and follow the
time evolution.
A calculated time evolution of
the density distrbituion is shown
in Fig. \ref{fig1}b.
The soliton is indeed reflected by the electron beam.
However, the solitonic character is not preserved
in that the wave packet quickly spreads out after reflection.
A fraction of 0.54 of the initial condensate is reflected,
other atoms are lost due to dissipation from the beam.
In Fig. \ref{fig1}c, a stronger dissipation strength parameter
$\gamma$
is used while all other parameters are kept the same. The reflection fraction rises from 0.54
to 0.62. The wave packets spreads out slower after being reflected.

A very effective way to further increase the reflected fraction and
minimize the atom loss
is using an electron beam with a sharper edge.
This is because while a strong enough dissipation
will reflect atoms through QZD and cause very small
atom loss, a weaker dissipation is less
effective in reflecting and the soliton
would go through such weakly dissipative region, losing atoms.
However, before reaching the region where
QZD reflection happens effectively,
the soliton always has to go through
a region with weaker dissipation which causes atom loss.
By using a sharper beam edge, this lossy region
is greatly shortened, and as a result the atom
loss can be greatly reduced.

Fig. \ref{fig1}d shows the results for an electron beam with
a sharper edge with $w=0.1$ compared to $w=0.8$ in
Fig. \ref{fig1}c, while all other parameters are kept the same.
The reflected fraction rises greatly from 0.62 to 0.91.
Most importantly,
now the solitonic character is indeed preserved by the reflection
in that
the wave packet maintains a constant shape as it
propagates after the reflection.
This indicates the balance between dispersion and nonlinearity
required for shape stability is preserved
by the QZD reflection, even though a small fraction
of the atoms are lost due to dissipation.
In calculating Fig. \ref{fig1} the
initial soliton velocity $v=-0.45553$,
the initial position of the soliton $x_0=10$,
and the center position of the beam $x_\textrm{b}=-5$.

The principle quantity of interest here is the successfully
reflected fraction
$P_{\textrm{refl}} \equiv \int_{x_\textrm{b}}^{\infty}
|\psi(x,t_{\textrm{final}})|^2 dx$,
where $t_{\textrm{final}}$ is the end time
of the evolution chosen such that
the reflection has been completed and no further atom loss happens.
On the one hand, a larger $P_{\textrm{refl}}$ means
smaller atom loss according to its definition.
On the other hand, according to our experience, the higher
the $P_{\textrm{refl}}$ value, the better the solitonic character
is preserved after reflection.

Apart from the sharpness of the beam edge, other factors can also
influence the value of $P_{\textrm{refl}}$.
These include the strength parameter of the dissipation $\gamma$,
and the velocity $v$ of the incoming soliton. As one would naturally
expect for QZD, for stronger dissipation, due to QZD
the boundary of the monitored region will
be more similar to an impenetrable wall.
As a result the atom loss during reflection will be
reduced. For the velocity dependence, the essence of QZD is that
the coherences between the monitored and unmonitored regions
are repeatedly set to zero by the measurements with
a high enough repetition frequency that coherent transmission
between the regions cannot build up.
Whether the repetition frequency is high enough
is affected by the kinetic energy. If the kinetic
energy is very high, the time scale of the coherent evolutions will
be very short so that the requirement for the repetition
frequency will be very high. From another point of view,
for lower kinetic energies, the same repetition frequency,
or equivalently the same intensity of a continuous monitoring
such as by the electron beam, can induce a more pronounced
QZD than for higher kinetic energies. So
for a slower incoming soliton, due to its lower kinetic energy
the QZD reflection will be more effective and the atom loss
will be reduced.

In Figs. \ref{fig2}a-f we show
the results of a systematic investigation
on the effects of varying $w$, $v$ and $\gamma$.
As can be seen, $P_{\textrm{refl}}$ is higher when the sharpness
parameter $w$ decreases, or when
the incoming velocity $v$ decreases, or when the strength
parameter of dissipation $\gamma$ increases.
In the overwhelming majority of cases,
only reflection is visible.
For high incoming velocity, small dissipation
and a narrow beam, a
faint transmission is also visible, as shown in
Fig. \ref{fig2}g.

For very small incoming velocity, the reflected
fraction $P_{\textrm{refl}}$
can be very high. In Fig. \ref{fig2}h such
a case is shown. The incoming velocity $v=-0.015625$.
$P_{\textrm{refl}}$ reaches a very high value of 0.996
and the solitonic character is very well preserved.
Curiously, as can be seen from the figure,
for such small incoming velocity
of the soliton, even at the closest
encounter there is
a pronounced gap between the beam and
the soliton. The electron
beam seemingly can reflect the soliton without any contact,
an impossible feat.
In Fig. \ref{fig2}i we plot the
electron beam profile $\Gamma(x)$
together with the density distribution $|\psi(x)|^2$
and the amplitude distribution $|\psi(x)|$
of the soliton at the closest encounter at $t=640$.
As shown, although the density
distribution seems quite cleanly separated from the electron
beam, there is a sizable presence of the amplitude
near the position of the beam. This close encounter
effects the reflection.

With the numerical evidences presented so far for the
beneficiary effect of having a small $w$ for promoting
$P_{\textrm{refl}}$, one may still ask the question,
is a small $w$ advantageous because it describes
a sharper beam edge or a narrower beam width?
Considering numeric evidences presented so far these two
views are indistinguishable,
because a small $w$ means both a sharper edge and a
narrower beam width, due to the Gaussian form we have
chosen for the beam (cf. equation (\ref{Gaussianbeam})).
To have a numeric resolution of this question,
we investigate a beam profile given by
\begin{eqnarray}
\Gamma_{\textrm{new}}(x)=\left\{\begin{matrix}
e^{-(\frac{x-x_\textrm{l}}{w})^2}, & x<x_\textrm{l}\\
1, & x_\textrm{l} \leq x < x_\textrm{r}\\
e^{-(\frac{x-x_\textrm{r}}{w})^2}, & x \geq x_\textrm{r}.
\end{matrix}\right.
\end{eqnarray}
This profile is shown by the black solid
line in Fig. \ref{fig2}j.
This new beam has the same sharp
edge as a Guassian beam with the same $w=0.1$ (red dashed curve
in Fig. \ref{fig2}j),
but is much wider.
These two beam profile both give $P_{\textrm{refl}}=0.94$,
while a Gaussian beam with similar width to the new beam
(blue dotted curve in Fig. \ref{fig2}j with $w=0.8$)
gives $P_{\textrm{refl}}=0.76$.
Since the new beam has a sharp edge but not a narrow width,
this evidences that it is the sharpness
of the beam edge instead of a narrow beam width
that is beneficiary in promoting $P_{\textrm{refl}}$.

Other parameters used in calculating Fig. \ref{fig2} are:
$x_0=10$, $x_\textrm{b}=-5$ except for Figs. \ref{fig2}h and i
where $x_\textrm{b}=-7$.
Fig. \ref{fig2}a, $\gamma=400$, $v=-0.25$. Fig. \ref{fig2}b,
$w=0.1$, $v=-0.25$. Fig. \ref{fig2}c, $w=0.1$, $\gamma=400$.
Fig. \ref{fig2}d, $\gamma=400$. Fig. \ref{fig2}e, $v=-0.25$.
Fig. \ref{fig2}f, $w=0.1$.
Fig. \ref{fig2}g, $w=0.1$, $v=-1.51241$, $\gamma=25$.
Figs. \ref{fig2}h and i, $\gamma=100$, $w=0.1$.
Fig. \ref{fig2}j,
$x_{\textrm{l}}=-6$, $x_{\textrm{r}}=-5$,
$v=-0.25$, $\gamma=200$.

\textbf{An example of application}. As an example of application for the above discussed QZD
reflection of a bright soliton by
an electron beam, we show that the translational
motion of a bright soliton can be precisely
stopped using this mechanism.
As seen in Figs. \ref{fig1}b and \ref{fig1}c, and
Figs. \ref{fig2}g and \ref{fig2}h,
the reflected soliton always has the same speed as
that of the incoming
soliton. This inspires us to learn from a soccer
player to use the mechanism in the following
way in order to stop the soliton:
we employ a moving electron beam that is moving in the same
direction of the soliton and with half the soliton's velocity,
cf. Fig. \ref{fig3}a.
The beam profile now is given by the time-dependent function:
\begin{eqnarray}
\Gamma_{\textrm{mov}}(x,t)=e^{-\left [\frac{x-
(x_\textrm{b}- v t/2)}{w} \right]^2}.
\end{eqnarray}
As can be seen from the calculated time-evolution of the
density distribution shown in Fig. \ref{fig3}b,
after the beam-soliton encounter the soliton is
precisely stopped. The parameter used in calculating
Fig. \ref{fig3}b are $v=-0.25$,
$\gamma=100$, $w=0.1$, $x_\textrm{d}=-5$, $x_0=10$.
The same method can also be employed to engineer
the final velocity of the soliton to be other chosen values.

\textbf{Discussion}

In summary, we have shown that the QZD induced
by an electron beam can reflect a bright soliton
supported by an attractive Bose-Einstein condensate
and the solitonic character can be well preserved.
This reflection at the macroscopic scale constitute
a demonstration and test of the
subtle and peculiar character of the quantum measurement process.
According to our best knowledge, this idea of reflecting a bright soliton using
pure dissipation has not been put forward before.
Technically, the key to
preserve the solitonic character and minimize atom loss
is to have a sharp beam edge, apart from
the perhaps more naturally expected high dissipation strength.
Also, if one can choose to work with a slow incoming soliton,
the performance will also improve dramatically.
Finally, we have shown that this phenomenon can be employed as an entirely new and
useful tool for manipulating the motional state of the
bright soliton.
Compared to the more familiar optical potentials generated by laser
beams, an electron beam can be focused much more sharply
and thus can explore regimes difficult
for an optical potential to reach.
Also, if used in combination with optical potentials,
it is readily conceivable that much more interesting
phenomena and techniques await to be discovered.
The phenomena studied in the present work
can be readily realized
using the recent experimental capability \cite{bar2013}
combining an electron beam and a Bose-Einstein condensate.
Our work shows this new experimental capability
offers a great opportunity to study the spatial QZD,
and more generally a many-body system out of equilibrium.

\maketitle

\clearpage
\newpage

\,

\,

\,

\,

\begin{acknowledgments}
The work is supported by the National Natural Science
Foundation of China (Grant No. 11575082, No. 11761161001,
No. 11535004, No. 11375086, and No. 11120101005, No. 11235001) and by
the International Science \& Technology Cooperation Program
of China (Grant No. 2016YFE0129300).
\end{acknowledgments}

\section*{Author contributions}
X.Z. performed the calculations, X.Z., X.F., C.X., Z.R. and J.P. discussed the results and wrote the paper.

\textbf{Competing interests:} The authors declare no competing interests.


\begin{thebibliography}{}
\bibitem{bru}Brukner, C. in Quantum [Un] Speakables II     95-117 (Springer, 2017).
\bibitem{leg}Leggett, A. The quantum measurement problem. science 307, 871-872 (2005).
\bibitem{fac2008}Facchi, P. \& Pascazio, S. Quantum Zeno dynamics: mathematical and physical aspects. Journal of Physics A: Mathematical and Theoretical 41, 493001 (2008).
\bibitem{fac2002}Facchi, P. \& Pascazio, S. Quantum zeno subspaces. Physical review letters 89, 080401 (2002).
\bibitem{hom}Home, D. \& Whitaker, M. A conceptual analysis of quantum Zeno; paradox, measurement, and experiment. Annals of Physics 258, 237-285 (1997).
\bibitem{sch}Schafer, F. et al. Experimental realization of quantum zeno dynamics. Nature communications 5 (2014).
\bibitem{sig}Signoles, A. et al. Confined quantum Zeno dynamics of a watched atomic arrow. Nature Physics 10, 715-719 (2014).
\bibitem{bar}Barontini, G., Hohmann, L., Haas, F., Esteve, J. \& Reichel, J. Deterministic generation of multiparticle entanglement by quantum Zeno dynamics. Science 349, 1317-1321 (2015).
\bibitem{bre}Bretheau, L., Campagne-Ibarcq, P., Flurin, E., Mallet, F. \& Huard, B. Quantum dynamics of an electromagnetic mode that cannot contain N photons. Science 348, 776-779 (2015).
\bibitem{kal}Kalb, N. et al. Experimental creation of quantum Zeno subspaces by repeated multi-spin projections in diamond. Nature communications 7, 13111 (2016).
\bibitem{wan}Wang, X.-B., You, J. \& Nori, F. Quantum entanglement via two-qubit quantum Zeno dynamics. Physical Review A 77, 062339 (2008).
\bibitem{man}Maniscalco, S., Francica, F., Zaffino, R. L., Gullo, N. L. \& Plastina, F. Protecting entanglement via the quantum Zeno effect. Physical review letters 100, 090503 (2008).
\bibitem{ros}Rossi, R., Romero, K. F. \& Nemes, M. Semiclassical dynamics from Zeno-like measurements. Physics Letters A 374, 158-160 (2009).
\bibitem{sha}Shao, X.-Q., Chen, L., Zhang, S. \& Yeon, K.-H. Fast CNOT gate via quantum Zeno dynamics. Journal of Physics B: Atomic, Molecular and Optical Physics 42, 165507 (2009).
\bibitem{cha}Chandrashekar, C. Zeno subspace in quantum-walk dynamics. Physical Review A 82, 052108 (2010).
\bibitem{shi}Shi, Z., Xia, Y., Wu, H. \& Song, J. One-step preparation of three-particle Greenberger-Horne-Zeilinger state via quantum Zeno dynamics. The European Physical Journal D 66, 127 (2012).
\bibitem{sha2017}Shao, X., Wu, J., Yi, X. \& Long, G.-L. Dissipative preparation of steady Greenberger-Horne-Zeilinger states for Rydberg atoms with quantum Zeno dynamics. Physical Review A 96, 062315 (2017).
\bibitem{bar2013}Barontini, G. et al. Controlling the dynamics of an open many-body quantum system with localized dissipation. Physical review letters 110, 035302 (2013).
\bibitem{bra2009}Brazhnyi, V. A., Konotop, V. V., Perez-Garcia, V. M. \& Ott, H. Dissipation-induced coherent structures in Bose-Einstein condensates. Physical review letters 102, 144101 (2009).
\bibitem{zez2012}Zezyulin, D., Konotop, V., Barontini, G. \& Ott, H. Macroscopic Zeno effect and stationary flows in nonlinear waveguides with localized dissipation. Physical review letters 109, 020405 (2012).
\bibitem{gli}Gligoric, G., Maluckov, A., Stepic, M., Hadzievski, L. \& Malomed, B. A. Two-dimensional discrete solitons in dipolar Bose-Einstein condensates. Physical Review A 81, 013633 (2010).
\bibitem{ped}Pedri, P. \& Santos, L. Two-dimensional bright solitons in dipolar Bose-Einstein condensates. Physical review letters 95, 200404 (2005).
\bibitem{str}Strecker, K. E., Partridge, G. B., Truscott, A. G. \& Hulet, R. G. Formation and propagation of matter-wave soliton trains. Nature 417, 150-153 (2002).
\bibitem{den}Denschlag, J. et al. Generating solitons by phase engineering of a Bose-Einstein condensate. Science 287, 97-101 (2000).
\bibitem{bur}Burger, S. et al. Dark solitons in Bose-Einstein condensates. Physical Review Letters 83, 5198 (1999).
\bibitem{bec}Becker, C. et al. Oscillations and interactions of dark and dark¨Cbright solitons in Bose¨CEinstein condensates. Nature Physics 4, 496-501 (2008).
\bibitem{and}Anderson, B. et al. Watching dark solitons decay into vortex rings in a Bose-Einstein condensate. Physical Review Letters 86, 2926 (2001).
\bibitem{eie}Eiermann, B. et al. Bright Bose-Einstein gap solitons of atoms with repulsive interaction. Physical review letters 92, 230401 (2004).
\bibitem{kha}Khaykovich, L. et al. Formation of a matter-wave bright soliton. Science 296, 1290-1293 (2002).
\bibitem{ngu}Nguyen, J. H., Luo, D. \& Hulet, R. G. Formation of matter-wave soliton trains by modulational instability. Science 356, 422-426 (2017).
\bibitem{mey}Meyer, N. et al. Observation of Two-Dimensional Localized Jones-Roberts Solitons in Bose-Einstein Condensates. Physical review letters 119, 150403 (2017).
\bibitem{mcd}McDonald, G. D. et al. Bright solitonic matter-wave interferometer. Physical review letters 113, 013002 (2014).
\bibitem{hel}Helm, J., Rooney, S., Weiss, C. \& Gardiner, S. Splitting bright matter-wave solitons on narrow potential barriers: Quantum to classical transition and applications to interferometry. Physical Review A 89, 033610 (2014).
\bibitem{ger}Gertjerenken, B., Wiles, T. P. \& Weiss, C. Progress towards quantum-enhanced interferometry with harmonically trapped quantum matter-wave bright solitons. Physical Review A 94, 053638 (2016).
\bibitem{hel2012}Helm, J. L., Billam, T. P. \& Gardiner, S. A. Bright matter-wave soliton collisions at narrow barriers. Physical Review A 85, 053621 (2012).
\bibitem{mar}Marchant, A. et al. Controlled formation and reflection of a bright solitary matter-wave. Nature communications 4, 1865 (2013).
\end{thebibliography}
\end{document}